\input harvmac
\let\includefigures=\iftrue
\let\useblackboard==\iftrue
\newfam\black

\includefigures
\message{If you do not have epsf.tex (to include figures),}
\message{change the option at the top of the tex file.}
\input epsf
\def\figin{\epsfcheck\figin}\def\figins{\epsfcheck\figins}
\def\epsfcheck{\ifx\epsfbox\UnDeFiNeD
\message{(NO epsf.tex, FIGURES WILL BE IGNORED)}
\gdef\figin##1{\vskip2in}\gdef\figins##1{\hskip.5in}
\else\message{(FIGURES WILL BE INCLUDED)}%
\gdef\figin##1{##1}\gdef\figins##1{##1}\fi}
\def\DefWarn#1{}
\def\figinsert{\goodbreak\midinsert}
\def\ifig#1#2#3{\DefWarn#1\xdef#1{fig.~\the\figno}
\writedef{#1\leftbracket fig.\noexpand~\the\figno}%
\figinsert\figin{\centerline{#3}}\medskip\centerline{\vbox{
\baselineskip12pt\advance\hsize by -1truein
\noindent\footnotefont{\bf Fig.~\the\figno:} #2}}
\endinsert\global\advance\figno by1}
\else
\def\ifig#1#2#3{\xdef#1{fig.~\the\figno}
\writedef{#1\leftbracket fig.\noexpand~\the\figno}%
\global\advance\figno by1} \fi

\def\id{{1 \kern-.28em {\rm l}}}

\def\K3{{\bf K3}}
\def\journal#1&#2(#3){\unskip, \sl #1\ \bf #2 \rm(19#3) }
\def\andjournal#1&#2(#3){\sl #1~\bf #2 \rm (19#3) }

\def\bar{\overline}

\def\ie{{\it i.e.}}
\def\eg{{\it e.g.}}

\def\frac#1#2{{#1\over#2}}

\def\inbar{\,\vrule height1.5ex width.4pt depth0pt}
\def\IC{\relax\hbox{$\inbar\kern-.3em{\rm C}$}}
\def\IR{\relax{\rm I\kern-.18em R}}
\def\IP{\relax{\rm I\kern-.18em P}}

%
%

%
\catcode`\@=11
\def\slash#1{\mathord{\mathpalette\c@ncel{#1}}}
\overfullrule=0pt

\def\ZZ{{\cal Z}}

\def\underrel#1\over#2{\mathrel{\mathop{\kern\z@#1}\limits_{#2}}}

\catcode`\@=12


%

\def \sinh{{\rm sinh}}
\def \cosh{{\rm cosh}}


\def\ie{{\it i.e.}}
\def\eg{{\it e.g.}}


\lref\MAsratT{
M.~Asrat,
``Moving holographic boundaries,''
Nucl. Phys. B {\bf 1008}, 116699 (2024)
doi:10.1016/j.nuclphysb.2024.116699
[arXiv:2305.15744 [hep-th]].
}

\lref\MAsratTT{
M.~Asrat,
``Kalb-Ramond field, black holes and black strings in (2 + 1)D,''
JHEP {\bf 08}, 135 (2025)
doi:10.1007/JHEP08(2025)135
[arXiv:2410.07580 [hep-th]].
}

\lref\HawkingP{
S.~W.~Hawking and D.~N.~Page,
``Thermodynamics of Black Holes in anti-De Sitter Space,''
Commun. Math. Phys. {\bf 87}, 577 (1983)
doi:10.1007/BF01208266
}

\lref\WittenP{
E.~Witten,
``Anti-de Sitter space, thermal phase transition, and confinement in gauge theories,''
Adv. Theor. Math. Phys. {\bf 2}, 505-532 (1998)
doi:10.4310/ATMP.1998.v2.n3.a3
[arXiv:hep-th/9803131 [hep-th]].
}

\lref\BerkoozP{
M.~Berkooz, Z.~Komargodski and D.~Reichmann,
``Thermal AdS(3), BTZ and competing winding modes condensation,''
JHEP {\bf 12}, 020 (2007)
doi:10.1088/1126-6708/2007/12/020
[arXiv:0706.0610 [hep-th]].
}

\lref\AdamsP{
A.~Adams, J.~Polchinski and E.~Silverstein,
``Don't panic! Closed string tachyons in ALE space-times,''
JHEP {\bf 10}, 029 (2001)
doi:10.1088/1126-6708/2001/10/029
[arXiv:hep-th/0108075 [hep-th]].
}

\lref\AtickP{
J.~J.~Atick and E.~Witten,
``The Hagedorn Transition and the Number of Degrees of Freedom of String Theory,''
Nucl. Phys. B {\bf 310}, 291-334 (1988)
doi:10.1016/0550-3213(88)90151-4
}

\lref\HagedornP{
R.~Hagedorn,
``Statistical thermodynamics of strong interactions at high-energies,''
Nuovo Cim. Suppl. {\bf 3}, 147-186 (1965)
CERN-TH-520.
}

\lref\NishiokaP{
T.~Nishioka and T.~Takayanagi,
``AdS Bubbles, Entropy and Closed String Tachyons,''
JHEP {\bf 01}, 090 (2007)
doi:10.1088/1126-6708/2007/01/090
[arXiv:hep-th/0611035 [hep-th]].
}

\lref\KlebanovP{
I.~R.~Klebanov, D.~Kutasov and A.~Murugan,
``Entanglement as a probe of confinement,''
Nucl. Phys. B {\bf 796}, 274-293 (2008)
doi:10.1016/j.nuclphysb.2007.12.017
[arXiv:0709.2140 [hep-th]].
}

\lref\AsratP{
M.~Asrat,
``Entropic $c$-functions in $T{\bar T}, J{\bar T}, T{\bar J}$ deformations,''
Nucl. Phys. B {\bf 960}, 115186 (2020)
doi:10.1016/j.nuclphysb.2020.115186
[arXiv:1911.04618 [hep-th]].
}

\lref\BahP{
I.~Bah, A.~Faraggi, L.~A.~Pando Zayas and C.~A.~Terrero-Escalante,
``Holographic entanglement entropy and phase transitions at finite temperature,''
Int. J. Mod. Phys. A {\bf 24}, 2703-2728 (2009)
doi:10.1142/S0217751X0904542X
[arXiv:0710.5483 [hep-th]].
}

\lref\BanadosP{
M.~Banados, C.~Teitelboim and J.~Zanelli,
``The Black hole in three-dimensional space-time,''
Phys. Rev. Lett. {\bf 69}, 1849-1851 (1992)
doi:10.1103/PhysRevLett.69.1849
[arXiv:hep-th/9204099 [hep-th]].
}

\lref\BanadosPP{
M.~Banados, M.~Henneaux, C.~Teitelboim and J.~Zanelli,
``Geometry of the (2+1) black hole,''
Phys. Rev. D {\bf 48}, 1506-1525 (1993)
[erratum: Phys. Rev. D {\bf 88}, 069902 (2013)]
doi:10.1103/PhysRevD.48.1506
[arXiv:gr-qc/9302012 [gr-qc]].
}

\lref\CasiniP{
H.~Casini and M.~Huerta,
``A c-theorem for the entanglement entropy,''
J. Phys. A {\bf 40}, 7031-7036 (2007)
doi:10.1088/1751-8113/40/25/S57
[arXiv:cond-mat/0610375 [cond-mat]].
}

\lref\RyuP{
S.~Ryu and T.~Takayanagi,
``Holographic derivation of entanglement entropy from AdS/CFT,''
Phys. Rev. Lett. {\bf 96}, 181602 (2006)
doi:10.1103/PhysRevLett.96.181602
[arXiv:hep-th/0603001 [hep-th]].
}

\lref\RyuPP{
S.~Ryu and T.~Takayanagi,
``Aspects of Holographic Entanglement Entropy,''
JHEP {\bf 08}, 045 (2006)
doi:10.1088/1126-6708/2006/08/045
[arXiv:hep-th/0605073 [hep-th]].
}

\lref\LewkowyczP{
A.~Lewkowycz and J.~Maldacena,
``Generalized gravitational entropy,''
JHEP {\bf 08}, 090 (2013)
doi:10.1007/JHEP08(2013)090
[arXiv:1304.4926 [hep-th]].
}

\lref\AsratPP{
M.~Asrat and J.~Kudler-Flam,
``$T\bar{T}$, the entanglement wedge cross section, and the breakdown of the split property,''
Phys. Rev. D {\bf 102}, no.4, 045009 (2020)
doi:10.1103/PhysRevD.102.045009
[arXiv:2005.08972 [hep-th]].
}

\Title{
} {\vbox{
{\vbox{
\centerline{Thermal/Black hole phase transition and} }}
\smallskip
\centerline{ entanglement entropy in (2 + 1)D} }}

\bigskip
\centerline{\it Meseret Asrat}
\smallskip
\centerline{International Center for Theoretical Sciences}
\centerline{Tata Institute of Fundamental Research
} \centerline{Bengaluru, KA 560089, India}

\smallskip

\vglue .3cm

\bigskip

\let\includefigures=\iftrue
\bigskip
\noindent

We consider a one parameter family of holographic solutions in classical string theory in three spacetime dimensions. In Euclidean space, the solutions interpolate smoothly without developing a conical singularity between the cigar black hole times a (non contractible) spatial circle and a thermal solution which has a (non contractible) temporal circle. We study the phase transition and the holographic entanglement entropy.


\bigskip

\Date{09/25}

\newsec{Introduction}

In this paper, we consider a special case of the solutions obtained in \refs{\MAsratTT, \MAsratT}. In this special case the solutions are parameterized by a single number $\gamma$. The parameter $\gamma$ takes the values in the interval $[-1,1]$. It plays the role of temperature. At $\gamma = -1$, in Euclidean space, the solution is the cigar black hole times a spatial circle. The spatial circle is non contractible. At $\gamma = 1$, the solution has the topology of a cigar times a temporal circle. The time circle is non contractible and thus the solution is usually referred thermal. For the intermediate values of $\gamma$, the solutions interpolate smoothly without developing conical singularity. The special holographic solutions we study are asymptotically flat (\ie, the scalar curvature approaches asymptotically to zero)  with a linear dilaton field.

In five spacetime dimensions there exists a (first order) phase transition between AdS-Schwarzschild black hole and thermal AdS \refs{\HawkingP, \WittenP}. The transition is called the Hawking-Page transition. The low temperature (semi-classical) Euclidean gravitational path integral is dominated by thermal AdS. At high temperature, it is dominated by the (large) black hole solution. In the dual boundary gauge theory the phase transition corresponds to the confinement-deconfinement transition \WittenP.

There also exists phase transition in three spacetime dimensions between the (non-rotating) BTZ black hole and thermal $AdS_3$ \refs{\HawkingP, \WittenP}. The thermal AdS solution, however, is separated from the continuous black hole spectrum by a mass gap \refs{\BanadosP, \BanadosPP}. Therefore, the BTZ black hole and thermal AdS geometries do not interpolate smoothly. The geometries in the gap have conical singularities.\foot{String theory allows/accommodates conical singularities. They are resolved by including twisted sectors.} The solutions we consider are special because they are regular and, moreover, they are connected smoothly. They can be useful to better understand the different/various phase transitions: Atick-Witten, Hagedron and Hawking-Page phase transitions, and the dynamics of (the associated) tachyons winding modes \refs{\AtickP, \HagedornP, \HawkingP, \BerkoozP}. 

The paper is organized as follows. In section two we obtain the solutions of interest. In section three we study the thermal phase transition. Entanglement entropy can be an order parameter of phase transitions. Thus, in section three, we compute along the transition the entanglement entropy of an interval using holography. We also study the entropic c-function \CasiniP. Unlike the entanglement entropy, it is ultraviolet cut-off/regulator independent.\foot{However, in general, there is no (to the best of my knowledge) a geometrical object in the bulk that directly computes it. It is derived from the entanglement entropy. When the entanglement is linear in the interval length, it is given by the finite part.} We find that the entropic c-function is always positive, and it is either a continuous smooth or a piecewise smooth function. We show that the c-function exhibits a phase transition as $\gamma$ varies. We also find that it exhibits a phase transition as the interval length changes, and thus, more generally, it can be used as a probe of (the existence of) phase transition(s). In section four we discuss future research directions.

\newsec{The solutions}

The solutions we consider are particular examples of the black strings \MAsratTT. The black string solutions \MAsratTT\ are obtained from ${\cal A}_3$ \MAsratT\ by applying simple coordinate transformations/identifications \MAsratTT. We refer to \MAsratTT\ for details and a comprehensive discussion. The black string solutions (specified by the four parameters $(\rho_-, \rho_+; \gamma; \lambda)$) are
\eqn\mmmXT{\eqalign{
ds^2 & = \frac{l^2\rho^2d\rho^2}{(\rho^2 - \rho_+^2)(\rho^2 - \rho_-^2)} - e^{2\phi}\frac{(\rho^2 - \rho_+^2)(\rho^2 - \rho_-^2)}{l^2\rho^2}dt^2 + e^{2\phi}\rho^2\left(dx - \frac{\rho_+\rho_-}{l\rho^2}dt\right)^2,\cr
B & = B_{tx}dt\wedge dx, \cr
B_{tx} & = \frac{1}{2l}e^{2\phi}\left[\rho_+^2(1 + \lambda + \gamma(1 + 2\lambda) + \lambda\gamma^2) + \rho_-^2(1 - \lambda - \gamma(1 - 2\lambda) - \lambda\gamma^2) - 2\rho^2(1 + 2\lambda \gamma)\right],\cr
Q & = e^{-2\Phi} {\star H} = 2\left|\frac{1 - \gamma^2}{lg_s^2}\right|,\cr
e^{2\Phi} & = g_s^2 |e^{2\phi}|,
}
}
where
\eqn\oooXT{\eqalign{
e^{-2\phi} & = \frac{\rho_+^2 (1 + \gamma)^2 - \rho_-^2(1 - \gamma)^2 - 4\gamma \rho^2}{\rho_+^2 - \rho_-^2}.
}
}
$Q$ is the axion charge per unit length. $\gamma$ takes its value in the range $[-1,1]$. $\lambda$ is an arbitrary constant which determines the value of the Kalb-Ramond $B$ field, \ie\ $B_{tx}$, at $\rho = \infty$. It does not enter in $H$.\foot{The black string solution is invariant under the combined transformations $\rho_\pm \to \rho_\mp$, $\gamma \to -\gamma$ and $\lambda \to -\lambda$. Therefore, positive $\gamma$ solutions can be obtained from negative $\gamma$ solutions by simply making the changes $\rho_\pm \to \rho_\mp$ and $\lambda \to -\lambda$.} In this paper we set $\lambda = -1/2\gamma$.

We now specialize the solutions to the specific case in which $(\rho^2_-, \rho^2_+) = (0, -4\gamma)$. This gives
\eqn\newaaa{\eqalign{
ds^2 & = \frac{l^2d\rho^2}{\rho^2 + 4\gamma} - e^{2\phi}\left(\frac{\rho^2 + 4\gamma}{l^2}\right)dt^2 + e^{2\phi}\rho^2 dx^2,\cr
B_{tx} & = {1\over l}\left[1 - \gamma^2\over (1 + \gamma)^2 + \rho^2\right],\cr
e^{2\Phi} & = g_s^2e^{2\phi}, \quad e^{2\phi} = {1\over (1+\gamma)^2 + \rho^2}.
}
}
The solution is invariant under $\gamma \to 1/\gamma$, $\rho \to \gamma\rho$ and $g_s \to g_s/\gamma$ (and $x\to-x$). The surface gravity at the horizon ($\rho^2_+ = -4\gamma > 0$) is
\eqn\newaaaa{\kappa = {\sqrt{-4\gamma}\over l^2(1-\gamma)}.
}
It can be obtained by requiring the smoothness of the Euclidean metric. The thermodynamics temperature is $T = \kappa/2\pi$. In the Euclidean geometry, the (Euclidean) time coordinate $\tau = -it$ has period $1/T$. The solution has a conical singularity in the interval $0 < \gamma < 1$ unless the angle coordinate $x$ has period $2\pi l(1+\gamma)/\sqrt{4\gamma}$.\foot{Conical (ALE) singularities are harmless in string theory. They allow a consistent string propagation.}

A semi-classical analysis of the Euclidean path integral shows that at low temperature $T < T_H$ the thermal solution is dominant. At high temperature $T > T_H$ the black hole is the dominant solution. The Hawking-Page temperature is $T_H = 1/(2\pi l^2)$. This is easily shown using modular transformation.

We are interested in solutions that are regular and smooth as $\gamma$ varies. To this end we introduce a new coordinate 
\eqn\newaab{
\rho = (1+\gamma)\sinh\theta + (1 - \gamma)\cosh\theta.
}
The solution, upon rescaling $t$ by $l^2$ and $x$ by $l$, becomes
\eqn\newaac{\eqalign{
     ds^2 & = l^2\left[d\theta^2 -\frac{(\gamma + e^{2\theta})^2}{(1 + e^{2\theta})(\gamma^2 + e^{2\theta})}dt^2 + \frac{(\gamma - e^{2\theta})^2}{(1 + e^{2\theta})(\gamma^2 + e^{2\theta})}dx^2\right], \cr
       B_{tx} & = \frac{l^2(1-\gamma^2)e^{2\theta}}{(1 + e^{2\theta})(\gamma^2 + e^{2\theta})},\cr       
         e^{2\Phi} & = g_s^2e^{2\phi}, \quad e^{2\phi} = \frac{e^{2\theta}}{(1 + e^{2\theta})(\gamma^2 + e^{2\theta})}, \quad -1 \leq \gamma \leq 1.
}
}
We, for the most part, consider the Euclidean version of the theory. To study the thermal phase transition we compactify the Euclidean time $\tau \sim \tau + \beta$. $\theta$ is a radial coordinate and it is positive.\foot{The solution is invariant under $\gamma \to 1/\gamma, \theta \to \theta - (1/2)\ln\left(\gamma^2\right), g_s \to g_s/\gamma, x \to -x$.} For large $\theta$ the metric is asymptotically flat (\ie, the scalar curvature approaches to zero) and the dilaton $\Phi$ is linear in $\theta$. In the rest of the paper our goal is to study \newaac. To the best of my knowledge \newaac\ has not been reported elsewhere in the literature.

We now consider \newaac\ for some special values of $\gamma$. 

At $\gamma = 1$, we have $B_{\tau x} = 0$. The metric and dilaton reduce to
\eqn\newaacc{\eqalign{
    ds^2 & = l^2(d\theta^2 +  d\tau^2 + \tanh^2\theta dx^2),\cr
    e^{2\Phi} & = {g_s^2\over 4\cosh^2\theta}.
    }
}
$\tau$ is the Euclidean time and it has period $\beta$. In general $\beta$ is arbitrary. $\theta$ is the radial coordinate, \ie\ $\theta \geq 0$. $x$ has period $2\pi$. The metric has the topology of a solid torus with contractible spatial circle. The spatial section is a disk. The solution is related to thermal $AdS_3$ \refs{\MAsratT, \BerkoozP}. 

At $\gamma = 0$, the solution becomes
\eqn\newaad{\eqalign{
    ds^2 & =  l^2\left[d\theta^2 + {e^{2\theta}\over 1 + e^{2\theta}}\left(d\tau^2 +  dx^2\right)\right],\cr
     B_{\tau x} & = \frac{il^2}{(1 + e^{2\theta})},\cr   
    e^{2\Phi} & = {g_s^2 \over 1 + e^{2\theta}}.
    }
}
$B_{\tau x}$ is complex since it is real in the Lorentzian space. The solution has no conical singularity. The metric has a $\ZZ_2$ symmetry which exchanges the two circles. This solution is related to Poincar\'e  AdS. The spatial section at constant Euclidean time is an annulus \AdamsP. Thus, a puncture that widens as we lower $\gamma$ appears at the center of the disk.

At $\gamma = -1$, we have $B_{\tau x} = 0$. The metric and dilaton reduce to
\eqn\newaae{\eqalign{
    ds^2 & = l^2(d\theta^2 + \tanh^2\theta d\tau^2 + dx^2),\cr
    e^{2\Phi} & = {g_s^2\over 4\cosh^2\theta}.
    }
   } 
 The metric has the topology of a solid torus with contractible temporal circle. It is the two dimensional Euclidean black hole times a spatial circle. $t$ has period $2\pi$. $x$ has period $\beta$. The solution is related to the BTZ black hole.
 
 Thus, as we change $\gamma$ from $\gamma = 1$ to $\gamma = -1$, the topology of the spatial section changes from a disk to a cylinder. The metric \newaac\ is $\ZZ_2$ symmetric under the exchanges $\gamma \to -\gamma$ and $\tau \leftrightarrow x$. The process is understood in terms of condensation of (competing) tachyonic winding modes along the two circles. We refer to \BerkoozP\ for more details and a comprehensive discussion.
 
 We next study using holography the entanglement entropy and the entropic c-function for an interval in the dual boundary theory.
 
\newsec{Entanglement entropy}

 We now compute the entanglement entropy for a spatial interval of length $\delta L$ with endpoints at $x = 2\pi L_1/L$ and $x = 2\pi L_2/L$. $L$ is the size of the boundary spatial circle. In the context of the AdS/CFT correspondence, entanglement entropy is given by the area of a homologous co-dimension two minimal surface in the bulk geometry \refs{\RyuP\RyuPP-\LewkowyczP}. In general there are two surfaces: a continuous smooth and piecewise smooth surfaces. The entropy (in a large class of interesting theories) is always given by (the area of) the continuous surface whenever it exists \refs{\BahP}. In a situation where the continuous surface does not exist, the entropy is given by the piecewise smooth surface.  
 
The continuous static surface is parametrized by $\theta = \theta(x)$. The entanglement entropy is 
 \eqn\newaaaf{
S_C = {l\over 4G_N}\int dx  e^{-2\phi}\sqrt{(\partial_x \theta)^2 + {(\gamma - e^{2\theta})^2\over (1 + e^{2\theta})(\gamma^2 + e^{2\theta})}}.
 }
 We write this in the following form
 \eqn\newaaag{
    S = {l\over 4G_{N}}\int dx F^2\sqrt{(\partial_x\theta)^2 + G^2} = {l\over 4G_{N}}\int dx {\cal L},
}
where
 \eqn\newaaah{
    F^2 = {(1 + e^{2\theta})(\gamma^2 + e^{2\theta})\over e^{2\theta}}, \quad G^2 = {(\gamma - e^{2\theta})^2\over (1 + e^{2\theta})(\gamma^2 + e^{2\theta})}.
}
${\cal L}$ does not explicitly depend on $x$, and therefore
 \eqn\newaaai{\eqalign{
    {\cal H} & = {\partial {\cal L}\over \partial ({\partial_x}\theta)}(\partial_x \theta) - {\cal L},  \cr
     & = -{F^2G^2\over\sqrt{(\partial_x \theta)^2 + G^2}}, 
}
}
is a constant. We denote the minimum value of $\theta$ by $\theta_\star$. We are here assuming that $\theta_\star$, \ie, the continuous smooth curve, exists. At $\theta = \theta_\star$, $\partial_x \theta = 0$. Therefore,
 \eqn\newaaak{
    {d\theta\over dx} = \pm\sqrt{{F^4G^4\over F^4_\star G^2_\star} - G^2},
}
where $F_\star = F(\theta_\star)$ and $G_\star = G(\theta_\star)$. The length of the interval is related to $\theta_\star$. It is given by
 \eqn\newaaal{
 \delta L = {L\over 2\pi}\int_{2\pi L_1/L}^{2\pi L_2/L}dx = {L\over \pi}\int_{\theta_\star}^{\theta_\infty}{d\theta\over {\partial_x \theta}} = {L\over \pi}\int_{\theta_\star}^{\theta_\infty}d\theta {F^2_\star |G_\star|\over \sqrt{F^4G^4 - F^4_\star G^2_\star G^2}}.
}
The entanglement entropy is
 \eqn\newaaam{
    S_C = {2l\over 4G_N}\int_{\theta_\star}^{\theta_\infty}d\theta {F^2_\star |G_\star|\over \sqrt{F^4G^4 - F^4_\star G^2_\star G^2}} F^2 {F^2G^2\over F^2_\star |G_\star|} = {2l\over 4G_N}\int_{\theta_\star}^{\theta_\infty}d\theta {F^4 G^2\over \sqrt{F^4G^4 - F^4_\star G^2_\star G^2}}. 
}
where $\theta_\infty$ is an ultraviolet regulator.
 
The equation for the length $\delta L$ can be put into the form
\eqn\newaaan{
   \delta L = {L\over 2\pi}\sqrt{(y_\star + 1)(y_\star + \gamma^2)}\left({y_\star - \gamma \over y_\star}\right)\int_{y_\star}^{\infty}{dy\over y - \gamma}\sqrt{(y+1)(y + \gamma^2)\over (y - y_\star)(y - y_0)(y+y_+)(y+y_-)},
}
where
 \eqn\newaaao{\eqalign{
 y_\star & = e^{2\theta_{\star}}, \quad y_0 =  {\gamma^2\over y_\star},\cr
       y_\pm  & = {y_\star(1+\gamma^2) + (y_\star - \gamma)^2\over 2y_\star} \pm {\sqrt{(y_\star(1-\gamma)^2 + (y_\star - \gamma)^2)(y_\star(1+\gamma)^2 + (y_\star - \gamma)^2)}\over 2y_\star}.
}
} 
Similarly, the expression for the entropy can be put into the form
 \eqn\newaaaj{
    S_C = {l\over 4G_N}\int_{y_\star}^{y_\infty}dy{(y+1)(y+\gamma^2)(y-\gamma)\over y^2}\sqrt{(y + 1)(y + \gamma^2)\over (y-y_\star)(y - y_0)(y+y_+)(y+y_-)}.
}
In some cases there exists a maximum interval length $L_{\rm max}$ larger than which $y_\star$ does not exist. In these cases we consider the piecewise smooth curve. 

The entanglement entropy given by the piecewise smooth configuration is
\eqn\newaaaja{\eqalign{
S_P & =  {l\over 4G_N}\int_{1}^{y_\infty}dy {(1 +y)(\gamma^2 + y)\over y^2} + {l\over 4G_N}(2(\gamma^2 + 1))\sqrt{(\gamma - 1)^2\over 2(\gamma^2 + 1)}\cdot{2\pi \delta L\over L},\cr
&= {l\over 4G_N}\left[y_\infty + (1+\gamma^2)\ln(y_\infty)- 1 + \gamma^2 +  \sqrt{ 2(\gamma^2 + 1)(\gamma - 1)^2}\cdot{2\pi \delta L\over L} \right].
}
}
In general the entanglement entropy $S$ is given by the minimum of the two, \ie\ $S_C, S_P$.

An equally useful quantity is the entropic c-function which in any local quantum field theory is ultraviolet regulator independent and finite. It is derived from the entanglement entropy
\eqn\newaaap{C = \delta L{\partial S\over \partial \delta L}.
}
The entropic c-function is a useful quantity to probe phase transitions \refs{\NishiokaP, \KlebanovP}. 

In general it is hard to evaluate the integrals exactly. For the case $\gamma = 0$ we have a closed form for both the entropy and c-function \AsratP. In what follows we use numerics to evaluate the integrals. We set $l/4G_N = 1/6$.

For the range $\gamma \in (\gamma_0, 1)$ where $\gamma_0 \approx -0.5321$ the smooth surface does not exist larger than an interval of length $L_{\rm max}$, see Fig. 1. $L_{\rm max}$ depends on $\gamma$. It is given by \newaaan\ with $y_\star = 1$. It obeys the inequality $L_\star \leq L_{\rm max} < L/2$. The minimum $L_\star$ is at $ \gamma = \gamma_1$ where $\gamma_1 \approx 0.06425$. In this case, \ie\ $\gamma \in (\gamma_0, 1)$, for $\delta L \geq L_{\rm max}$ the piecewise surface gives the entanglement entropy. For $\gamma \leq \gamma_0$, $y_\star$ exists always and thus the smooth curve gives the entanglement entropy. We note, therefore, that for $\delta L < L_\star$ the entanglement entropy is given, independent of $\gamma$, by the continuous solution $S_C$.
\ifig\loc{The plot depicts the maximum interval length $L_{\rm max}$ per the size of the boundary circle $L$ as a function of $\gamma$. $L_\star \leq L_{\rm max} < L/2$. $L_{\rm max} = L_\star$ at $\gamma = \gamma_1 \approx 0.06425$. The orange line is $L_{\rm max}/L = 0.5$. The vertical line is $\gamma = \gamma_0 \approx -0.5321$. In the case $\delta L \geq L_{\rm max}$ the continuous surface ceases to exist for $\gamma \in (\gamma_0,1)$. For $\gamma \leq \gamma_0$, the continuous surface exists always independent of $\delta L$.}
{\epsfxsize3.8in\epsfbox{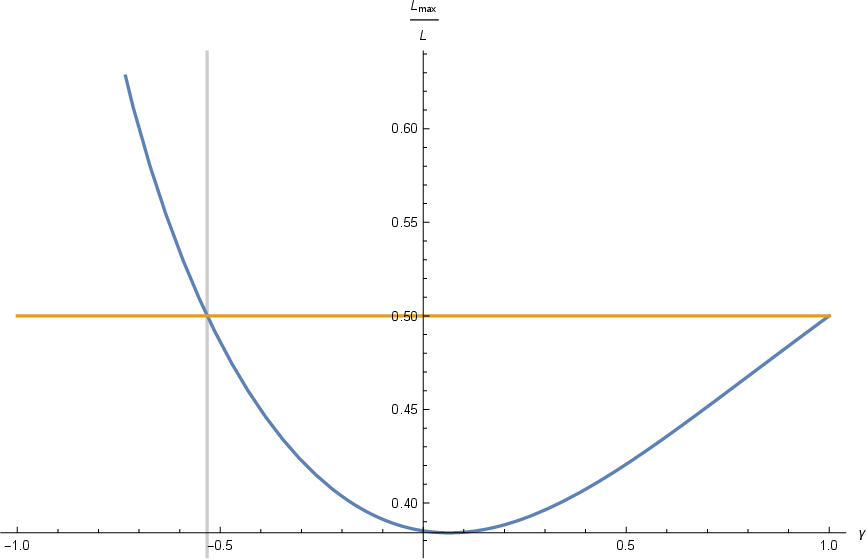}}

The entanglement entropy and the entropic c-function diverge as $\delta L$ approaches $L_0$, see \eg\ Fig. 2. $L_0$ is the minimum interval length that makes sense in the theory. It is the non-local scale of theory. It is given by
\eqn\newaaas{
{L_0\over L} = {1\over 4}.
}
We note that since the solution \newaac\ is independent of $\gamma$ for large $\theta$, the fact that $L_0$ is independent of $\gamma$ makes sense. 
\ifig\loc{The plot depicts the c-function as a function of the interval length $\delta L$ for $\gamma = 1$ and $\gamma = -1$. The blue curve is for $\gamma = 1$. The orange curve is for $\gamma = -1$. They diverge at $\delta L/L = 0.25$.}
{\epsfxsize5.5in\epsfbox{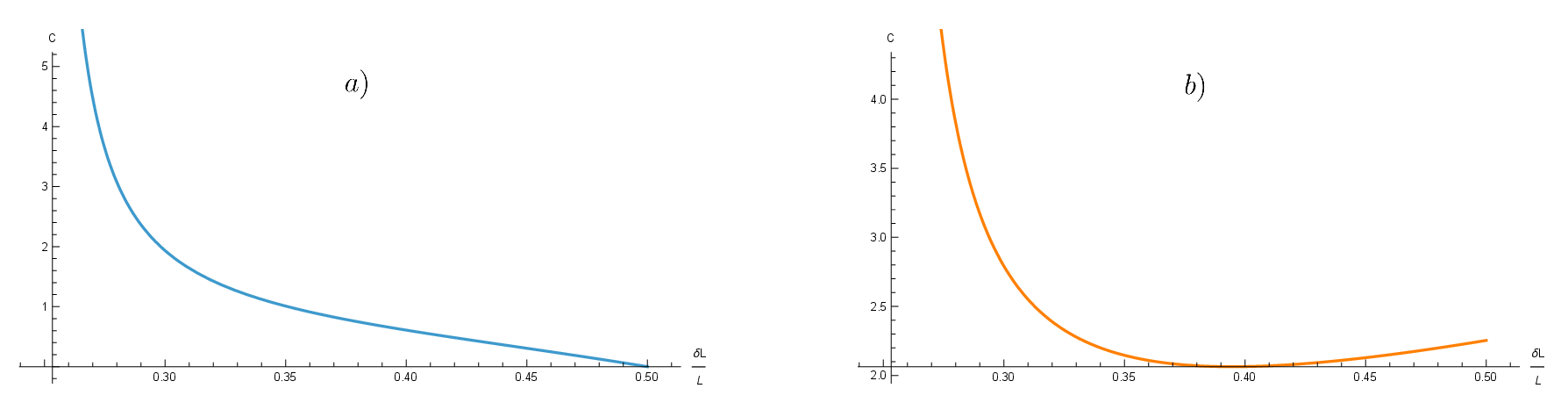}}

We now study the behavior of the entanglement entropy and therefore in turn also the entropic c-function as a function of the interval length $\delta L$. There are two cases two consider depending on the value of $\gamma$ relative to $\gamma_0$, see Fig 1. The first case is when $\gamma_0 < \gamma < 1$. The second case is when $\gamma \leq \gamma_0$. In the first case there is a maximum length $L_{\rm max}$ above which the continuous smooth surface ceases to exist. Therefore, for $\delta L \geq L_{\rm max}$ the entanglement entropy is given by the piecewise solution $S_P$ and the entropic c-function is linear in $\delta L$. However, in the second case since $y_\star$ exists always, the entanglement entropy is given by the continuous solution $S_C$.
 
In Fig. 3 we have a representative plot of the entropic c-function as a function of the interval length in the case where $\gamma_0 < \gamma < 1$. For definiteness, we have specifically chosen $\gamma = 0.4$. However, we would still have gotten a similar plot had we chosen any other $\gamma \in (\gamma_0, 1)$. The c-function exhibits a phase transition. It is piecewise smooth and continuous. Its slope changes sign at $L_{\rm max}$. The entanglement entropy is discontinuous. It makes a jump at $L_{\rm max}$. 
\ifig\loc{The plot depicts the c-function as a function of the interval length $\delta L$ for $\gamma = 0.4$. The c-function is normalized by $C_0 = C(L_{\rm max})$. Note $L_{\rm max}$ depends on $\gamma$. Therefore, in turn $C_0$ also depends on $\gamma$.}
{\epsfxsize3.5in\epsfbox{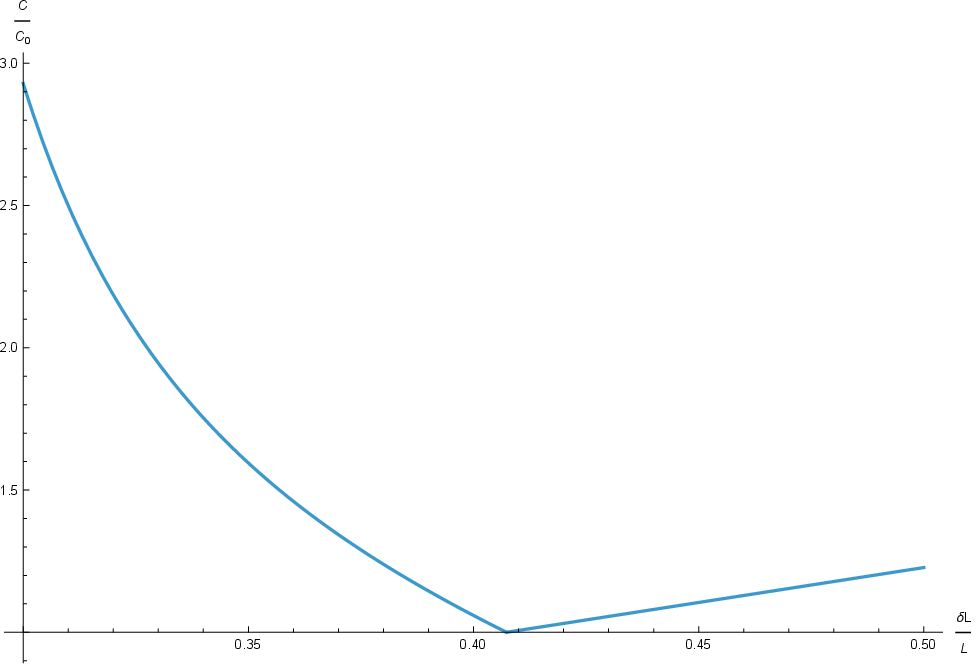}}

In Fig. 4 we have a representative plot of the entropic c-function as a function of the interval length for $\gamma \leq \gamma_0$. Here we chose $\gamma = -0.7$. In this case the c-function is smooth. However, here also its slope changes sign. See also plot $b)$ in Fig. 2. It is interesting to observe such a behavior since the minimal surface is given by the smooth solution \BahP. The entanglement entropy also exhibits a phase transition, \ie\ its slope changes sign, however it is not discontinuous.
\ifig\loc{The plot depicts the c-function as a function of the interval length $l$ for $\gamma = -0.7$. The c-function is normalized by $C_0(\gamma)$. $C_0$ is the minimum value that the c-function takes. In general it depends on $\gamma$.}
{\epsfxsize3.5in\epsfbox{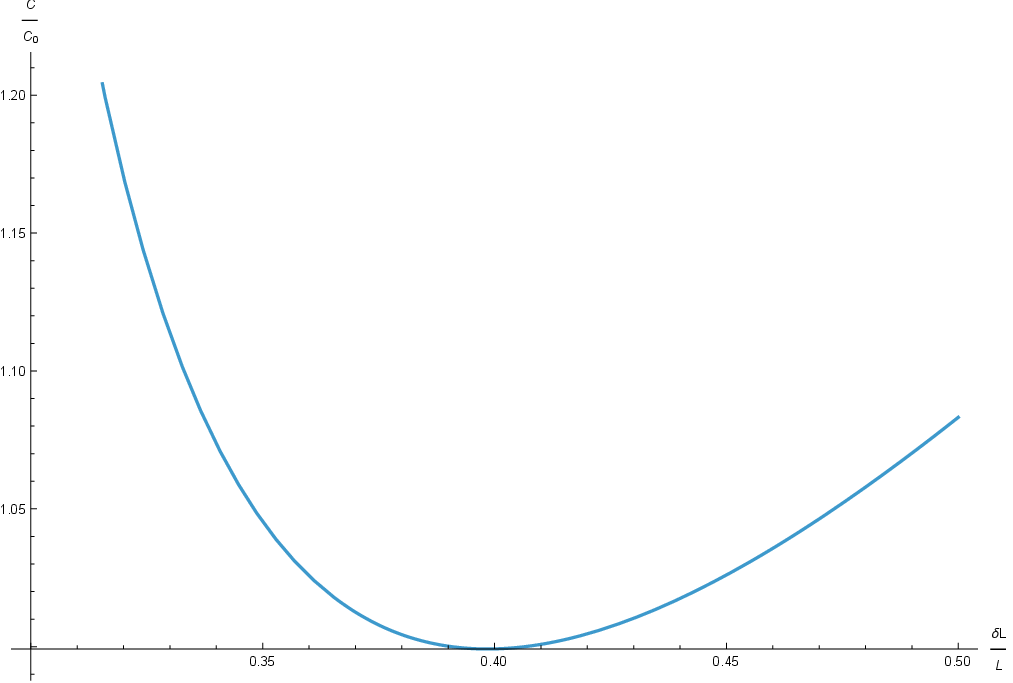}}

We next study the behavior of the entanglement and c-function as a function of $\gamma$. Here also we have two separate cases that we need to consider depending on the value of $\delta L$ in relation to $L_\star$, see Fig. 1.

We first consider the case $\delta L \leq L_\star$. In Fig. 5 we have a representative plot for the entropic c-function as a function of $\gamma$. In this case the entanglement entropy is given entirely by the continuous smooth surface configuration. However, we here also note that both the entanglement entropy and the entropic c-function exhibit a phase transition. Each of their slopes changes sign at a particular value of $\gamma$. However, both are continuous and smooth functions. 
\ifig\loc{The plot depicts the c-function as a function of $\gamma$ for interval length ${\delta L\over L} = 0.3$. The c-function is normalized with respect to $C_0$.}
{\epsfxsize3.5in\epsfbox{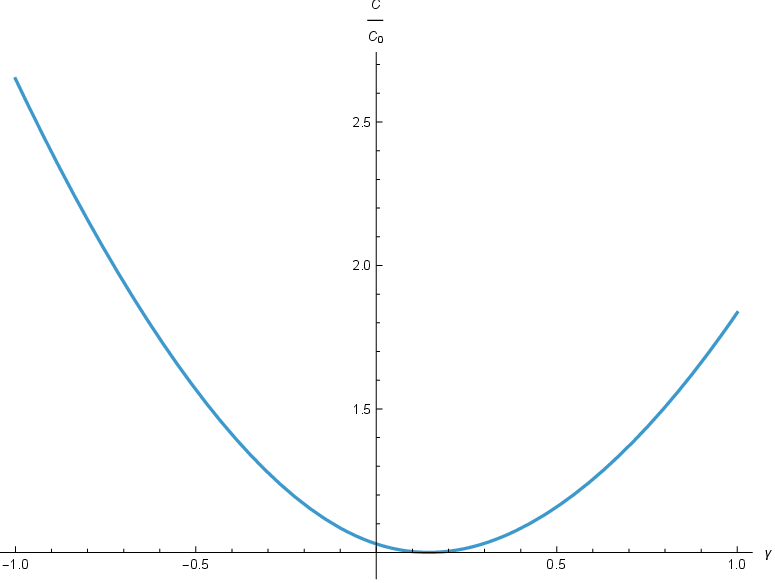}} 

The other case is $L_\star < \delta L < L/2$. In this case both the piecewise smooth and continuous smooth surfaces contribute. To give an example we take $\delta L/L = 0.45$. We note from Fig. 1 that there are two values of $\gamma$ for which $\delta L$ is $L_{\rm max}$. Let us denote these two values by $\gamma_-$ and $\gamma_+$. We assume $\gamma_- < \gamma_+$. In the example we are considering $\gamma_- \approx -0.4012$ and $\gamma_+ \approx 0.6910$. Then, it follows from Fig. 1 that for $\gamma < \gamma_-$ and $\gamma > \gamma_+$ the entanglement is given by $S_C$. For $\gamma_- \leq \gamma \leq \gamma_+$ however it is given by $S_P$. The entropic c-function is plotted in Fig. 6. We here see that the entropic c-function is piecewise smooth and continuous. It exhibits phase transitions. The entanglement entropy is discontinuous and makes jumps at $\gamma = \gamma_\pm$.
\ifig\loc{The plot depicts the c-function as a function of $\gamma$ for interval length ${\delta L\over L} = 0.45$.}
{\epsfxsize3.8in\epsfbox{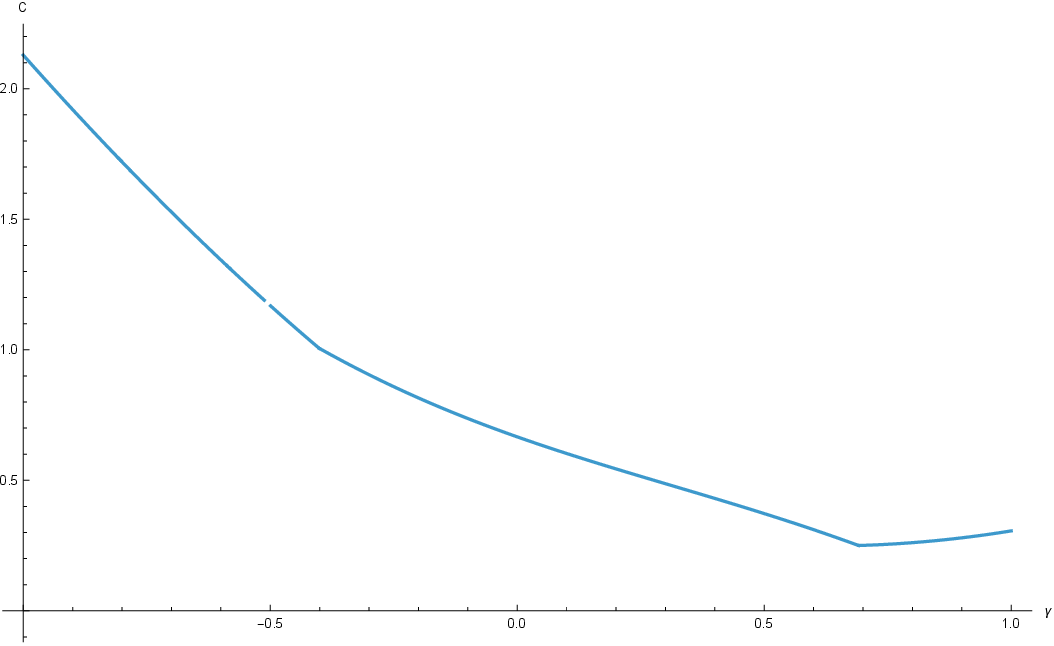}}

In the next section we summarize the main results and discuss future research directions.  

\newsec{Discussion}
 
In this paper we considered the one parameter family of holographic solutions \newaac\ which we again write below for convenience. They are a subclass of \MAsratTT.  After applying Wick rotation,\foot{It is obtained by making the analytic continuation $t \to i\tau$.} the solutions in Euclidean space are  
\eqn\newaaq{\eqalign{
     ds^2 & = l^2\left[d\theta^2 + \frac{(\gamma + e^{2\theta})^2}{(1 + e^{2\theta})(\gamma^2 + e^{2\theta})}d\tau^2 + \frac{(\gamma - e^{2\theta})^2}{(1 + e^{2\theta})(\gamma^2 + e^{2\theta})}dx^2\right], \cr
       B_{\tau x} & = \frac{i l^2(1-\gamma^2)e^{2\theta}}{(1 + e^{2\theta})(\gamma^2 + e^{2\theta})},\cr       
         e^{2\Phi} & = g_s^2e^{2\phi}, \quad e^{2\phi} = \frac{e^{2\theta}}{(1 + e^{2\theta})(\gamma^2 + e^{2\theta})}, \quad -1 \leq \gamma \leq 1.
}
}
$\tau$ is the Euclidean time and $x$ is an angle variable. $\theta$ is a radial coordinate. It is assumed to be positive. $\tau$ and $x$ are on a circle. Therefore, they are periodic. The metric has a $\ZZ_2$ symmetry. It exchanges $\tau \leftrightarrow x$ and $\gamma \to -\gamma$.

The solution at $\gamma = 1$ is related, through dualities,  to thermal $AdS_3$. Also the solution at $\gamma = -1$ is related to BTZ black hole. $\gamma$ plays the role of temperature. At high temperature, the black hole is the dominant solution. At low temperature the thermal solution dominates the action. The phase transition is understood in terms of competing winding modes \BerkoozP. The topology of the spatial section changes, as we increase $-\gamma$, from a disk to a cylinder. A widening puncture appears at the center of the disk. In holographic theories, the thermal phase transition is dual in the boundary theory to a confinement-deconfinement transition. 

Entanglement entropy is a useful quantity to identify and characterize phase transitions. The same information is contained in the ultraviolet cut-off/regulator independent quantity: the entropic c-function. It is derived from the entanglement entropy. In holographic theories the entanglement is related to a geometrical structure in the bulk. In this paper we computed, using holography, the entanglement entropy of an interval.  We studied the properties and/or behaviors of the entanglement entropy and the entropic c-function along the phase transition, \ie, along the topology changes. We also studied their dependence on the interval length.

We found, irrespective of $\gamma$, that both the entanglement and the c-function diverge as the interval length approaches the minimum value $L_0 = L/4$ where $L$ is the size of the boundary circle \newaaas.

We also noted that the entropic c-function is always positive and continuous. However, the entanglement entropy can be discontinuous and make jump(s). This is because in general there are two competing surfaces: a continuous smooth and piecewise smooth surfaces. Whenever the entanglement entropy is discontinuous, the entropic c-function is piecewise smooth.

We observed that, irrespective of the interval length, the c-function as a function of $\gamma$ changes its slope sign during the topology change process only once, see Fig. 5 and 6. However, the point at which it changes sign depends on the interval length. The c-function is piecewise smooth when the interval length is larger than $L_\star$, see Fig. 6. 

We also observed, for a given $\gamma \neq 1$, that the entropic c-function is a piecewise smooth function of the interval size only when $\gamma > \gamma_0$, see Fig. 3. For $\gamma \leq \gamma_0$, the c-function is a continuous smooth function, see Fig. 4. The point at which it changes its slope sign, in general, depends on $\gamma$. Therefore, more generally, the existence of a phase transition can be detected by studying the entropic c-function as a function of the size of the subsystem. 

An interesting research direction to explore is to use FZZ or related dualities to better understand the phase transition \ie, the dynamics of the winding modes along the two circles. It would be also useful to better understand the solution at $\gamma = 0$ \newaad\ but now with $\theta$ on the real line. In this case the two circles shrink to zero exponentially at large negative $\theta$. Other important direction to consider, which is of a particular interest, is to better understanding the physics at the non-local scale $L_0$ \AsratPP. I hope to investigate and address these and some other related points in a future work.

\bigskip\bigskip

\noindent{\bf Acknowledgements:} This work is supported by the Department of Atomic Energy under project no. RTI4001.

\listrefs

\bye